\documentclass[11pt]{article}
\usepackage{amssymb,amsmath,cite,geometry,graphicx,url,mathrsfs}
\newcommand{\dd}[1]{d^{#1}\mspace{-1.5mu}}
\catcode`\@=11

\@addtoreset{equation}{section}
\def\theequation{\arabic{section}.\arabic{equation}}
     
\catcode`\@=11
\def\thesection{\arabic{section}}

\def\appendix{\setcounter{section}{0}
        \def\thesection{Appendix.}
        \def\theequation{\Alph{section}.\arabic{equation}}}
\def\section{\@startsection{section}{1}{\z@}{3.5ex plus 1ex minus
   .2ex}{2.3ex plus .2ex}{\large\bf}}
     
\long\def\@makefntext#1{\parindent 0cm\noindent
\hbox to 1em{\hss$^{\@thefnmark}$}#1}

\newcommand{\captionfonts}{\small}
\makeatletter  % Allow the use of @ in command names
\long\def\@makecaption#1#2{%
  \vskip\abovecaptionskip
  \sbox\@tempboxa{{\captionfonts #1: #2}}%
  \ifdim \wd\@tempboxa >\hsize
    {\captionfonts #1: #2\par}
  \else
    \hbox to\hsize{\hfil\box\@tempboxa\hfil}%
  \fi
  \vskip\belowcaptionskip}
\makeatother   % Cancel the effect of \makeatletter  

\begin{document}
\begin{titlepage}
\vspace{.5in}
\begin{flushright}
October 2023\\  %date
revised December 2023\\
\end{flushright}
\vspace{.5in}
\begin{center}
{\Large\bf
 Covariant canonical quantization\\[.5ex] and the problem of time}\\  %title
\vspace{.4in}
{S.~C{\sc arlip}\footnote{\it email: carlip@physics.ucdavis.edu}\\
{W{\sc eixuan} H{\sc u}}\footnote{\it email: wxhu@ucdavis.edu}\\
       {\small\it Department of Physics}\\
       {\small\it University of California}\\
       {\small\it Davis, CA 95616}\\{\small\it USA}}
\end{center}

\vspace{.5in}
\begin{center}
{\large\bf Abstract}
\end{center}
\begin{center}
\begin{minipage}{4.75in}
{\small
In the covariant canonical approach to classical physics, each point 
in phase space represents an entire classical trajectory.  Initial data at a 
fixed time serve as coordinates for this ``timeless'' phase space, and 
time evolution can be viewed as a coordinate change.  We argue 
for a similar view in quantum theory.  As in the Heisenberg picture, 
the wave function is fundamentally time-independent.  On any given 
time slice, however, we can diagonalize a complete set of position
operators to form a basis, in which the projected wave function depends 
on the choice of time.  In this picture, time evolution can be viewed as 
a basis change in what is otherwise a block universe.  We argue that
this may help solve the ``problem of time'' in quantum gravity, and illustrate 
the idea with an example from three-dimensional quantum gravity.
}
\end{minipage}
\end{center}
\end{titlepage}
\addtocounter{footnote}{-2}

The ``problem of time'' \cite{Kuchar,Isham} is widely recognized as one of the 
fundamental conceptual obstacles to a quantum theory
of gravity.  Like other quantum theories, quantum gravity should presumably
describe the time evolution of quantum states and operators.  But conventional 
approaches to quantum evolution fail rather dramatically.  For a closed universe, 
for example, the standard time evolution operator, the Hamiltonian, is zero 
when acting on physical states, and the theory appears to be ``frozen.''   
For states that involve superpositions of sufficiently different physical configurations, 
it is not even clear that one can define a consistent ``time''  at all.

Time evolution is already subtle in classical general relativity.  As a dynamical 
theory, general relativity describes the evolution of spacetime geometry from some 
initial spatial hypersurface (or ``time slice'') to a later one.  But the theory is generally 
covariant, with no fixed background structure, so our choice of time coordinate 
can be changed arbitrarily without affecting the physics.  A time translation
$t\rightarrow t+\Delta t$ in coordinate time can be viewed as nothing more 
than a coordinate change.  Consistent with this picture, the Hamiltonian, which
generates such transformations, is  a constraint, vanishing on physical 
configurations.\footnote{In a spacetime with an asymptotic boundary, this is no 
longer quite true: the Hamiltonian acquires boundary terms, allowing a more traditional 
picture of time evolution, though only at the boundary.  This does not seem  relevant for our 
Universe, however, which appears to be asymptotically de Sitter, with no boundary.}

This does not prevent us from describing time evolution in classical general
relativity, of course.  In practice,  we do this by using some form of relational time.  
To describe the propagation of gravitational waves, for instance, we assume a fixed
background on which the waves are perturbations, and use that background to define 
a preferred time.  To describe our Solar System, we start with an approximation in 
which the Sun defines a preferred reference frame, and treat planetary motion as 
small perturbations.  To describe cosmology, we often use York's ``extrinsic time'' 
\cite{York}, in which the trace of the extrinsic curvature---physically, local Hubble expansion 
rate---serves as a preferred time.

In quantum gravity, though, relational time becomes much more difficult.  One
can measure time with GPS satellite signals \cite{RovelliGPS}, for instance,
or, reaching back in history, with the rotation of the Earth.  But what if one's wave 
function contains configurations with no GPS satellites, or no Earth?  One might
postulate a cloud of ideal ``clocks,'' as in \cite{BrownKuchar}, but these will back-react  
on the spacetime, changing the questions one is asking, and their quantum properties 
will make them imperfect time keepers \cite{UnruhWald,Pullin}.

We argue here that a partial  solution can be found in a somewhat unconventional view 
of classical mechanics.  The phase space of a classical theory, the starting point for 
canonical quantization, is usually described as the space of initial data on some time 
slice $\Sigma_t$, with additional structure (a symplectic form) determined by the details 
of the theory.   In the covariant phase space picture \cite{AshMag,Bombelli,Crn,Crnb,Lee,Waldb}, 
in contrast, each point in phase space is an entire classical history, that is, a complete 
solution of the classical equations of motion.  These two descriptions seem very different, 
but in a theory with a well posed initial value problem, they can be shown to be in 
one-to-one correspondence.

Despite this correspondence, though, the covariant phase space picture leads to 
a quantum theory that looks a bit unusual.  Quantization  automatically gives 
time-independent wave functions; in gravity, these are essentially amplitudes for 
entire block universes.  But by classically mapping the covariant phase space to 
the conventional phase space on a time slice $\Sigma_t$, we can obtain a set of
slice-dependent observables, which become ``time''-dependent operators in the 
quantum theory.  These operators, in turn, determine a ``time''-dependent 
basis for the Hilbert space, different for each time slice $\Sigma_t$.  The result is not
unlike the usual Heisenberg picture, but without any need for a preferred time coordinate.

We will elaborate on this picture below, arguing in particular that the initial data
on any time slice can serve as coordinates for the covariant phase space, with time 
evolution acting as a coordinate change.  After quantization,  such a coordinate change 
translates into a change of basis, where any classical choice of time slicing provides an 
admissible basis.  As in conventional quantum mechanics, we can change from a 
Heisenberg picture to a Schr{\"o}dinger picture by expressing wave functions in such a 
``time''-dependent basis.  But we will now have an infinite number of Schr{\"o}dinger pictures, 
one for each classical time slicing, all equivalent to the original Heisenberg picture.
We will give a simple example from gravity in three spacetime dimensions in which this 
program can be carried out in full, and will conclude with a discussion of the possibilities 
and difficulties of extending the approach to realistic quantum gravity.  

We do not claim mathematical rigor---for many of our arguments, rigorous treatments
are known only for simple cases---and we certainly do not pretend to have quantized
gravity.  Rather, our argument is that \emph{if} general relativity can be quantized using 
covariant phase space methods, this might simplify some of problems associated with
time in quantum gravity.

\section{The problem of time \label{time}}

We begin with a brief review of the problem(s) of time in quantum gravity.  This is a large
subject, and we refer readers to \cite{Kuchar,Isham} for more comprehensive treatments
and technical details.  We will focus on the most straightforward approach to quantum
gravity, Dirac quantization in the metric representation, but our conclusions carry over to other 
approaches such as the connection representation of loop quantum gravity \cite{loop}. 

Newtonian mechanics and nonrelativistic quantum mechanics assume an absolute time
that, in principle, provides a reference to describe time evolution.  Special relativity abandons 
absolute time, but still contains a collection of fixed, nondynamical reference frames that 
provide an unambiguous description of evolution.  Even quantum theory in a curved spacetime
has a fixed spacetime background to use as a reference for time evolution.

In general relativity, on the other hand, spacetime is dynamical, with no preferred specification 
of time.  The time coordinate $t$ is just that:  a human-made label with no intrinsic meaning, 
which can be chosen and changed arbitrarily.  This arbitrariness shows up in several ways:
\begin{enumerate}
\item {\bf Hamiltonian evolution:}\\[1ex]
The Hamiltonian description of a physical system requires a choice of time
coordinate, seemingly breaking general covariance.  A long effort to address this problem
culminated in the work of Arnowitt, Deser, and Misner (ADM) \cite{ADM}.  In the ADM
formalism, at time coordinate ``$t$'' is named, but is left completely unspecified except
for the requirement that the hypersurfaces of constant $t$ be spacelike.  A general
spacetime metric can be written as\footnote{We use the conventions of \cite{Carlip_GR},
which assume a metric with signature $(+---)$ , a choice that affects some signs.}
\begin{equation}
ds^2 = N^2dt^2 - q_{ij}(dx^i + N^idt)(dx^j + N^jdt) ,
\label{time1}
\end{equation}
where $q_{ij}$ is a positive definite spatial metric on each slice $\Sigma_t$ of constant $t$.
The Hamiltonian version of the Einstein-Hilbert action then takes the form
%yields a canonical momentum
%\begin{equation}
%\pi^{ij} = \frac{1}{2\kappa^2}\sqrt{q}(K^{ij} - q^{ij}K) ,
%\label{time2}
%\end{equation}
% where $\kappa^2=8\pi G$ is a constant and $K_{ij}$ is the extrinsic curvature of the slice $\Sigma_t$.
%\begin{equation}
%K_{ij} = -\frac{1}{2N}\left( \partial_0 q_{ij}   
%          - {}^{\scriptscriptstyle(3)}\nabla_iN_j - {}^{\scriptscriptstyle(3)}\nabla_jN_i\right)
%\label{time3}
%\end{equation}
% A long but unexciting calculation of the standard Einstein-Hilbert action then gives  
\begin{equation}
I = \int dt\int\dd{3}x\left[\pi^{ij}{\dot q}_{ij} - N\mathscr{H} - N_i\mathscr{H}^i\right] ,
\label{time4}
\end{equation}
where $\pi^{ij}$ is the momentum conjugate to $q_{ij}$ and
\begin{equation}
\mathscr{H}^i = -2\,{}^{\scriptscriptstyle(3)}\nabla_j\pi^{ij}  , \quad
\mathscr{H} = \frac{2\kappa^2}{\sqrt{q}}\left(\pi^{ij}\pi_{ij} - \frac{1}{2}\pi^2\right) 
    - \frac{1}{2\kappa^2}\sqrt{q}\,{}^{\scriptscriptstyle(3)}\!R  .
\label{time5}
\end{equation}
(Here $\kappa^2=8\pi G$ is a constant, ${}^{\scriptscriptstyle(3)}\nabla$ is the spatial covariant 
derivative compatible with  $q_{ij}$, $\pi = \pi^i{}_i$,  and ${}^{\scriptscriptstyle(3)}\!R$ is the intrinsic 
scalar curvature of the slice $\Sigma_t$).

The action (\ref{time4}) is a standard canonical action, and the Hamiltonian 
\begin{equation}
H = \int\dd{3}x (N\mathscr{H} + N_i\mathscr{H}^i)
\label{time5a}
\end{equation}
gives the correct Hamilton's equations of motion.   But the ``lapse function'' $N$ and 
the ``shift vector'' $N_i$ appear only as Lagrange multipliers, whose equations of motion 
are $H=0$.  Up to possible boundary terms, the Hamiltonian for general relativity is thus 
a \emph{constraint}, vanishing for physical configurations.  This reflects the fact that the 
time coordinate $t$ is arbitrary; a  translation in $t$ may look like time evolution, but it 
can be implemented by merely changing coordinates.

In classical general relativity, we have learned to live with this problem.  In a setting in 
which we have a time coordinate with a clear physical meaning---for instance, in most 
cosmological applications---we can think of the coordinate $t$ as being a physical clock time,
with evolution defined relative to this clock.  In quantum gravity, though, the 
situation is more complex.  The usual procedure for obtaining a Schr{\"o}dinger equation 
no longer gives us a time evolution equation, but rather the Wheeler-DeWitt equation \cite{DeWitt}
\begin{equation}
{\widehat H}|\psi\rangle = 0  .
\label{time6}
\end{equation}
The time derivative in the usual Schr{\"o}dinger picture has disappeared, leaving us with
a ``frozen'' formalism in which the physical states appear to be independent of time.   We
will see below that the same frozen time also occurs in the Heisenberg picture

Mimicking the classical procedure, we can attempt to extract a relational time from (\ref{time6}), 
but efforts in this direction have met with little success.  While there are natural choices 
of time for particular spacetimes, a wave function should allow a superposition of many 
geometries, and it is not at all clear how to extract a single relational time that makes sense for 
an arbitrary superposition.  For many years, it was widely hoped that York's ``extrinsic time''  
 \cite{York}, in which the mean extrinsic curvature $K$ serves as a time coordinate, might give 
 a universal slicing.  But we have now learned that an infinite family of spacetimes admit no 
 such hypersurfaces \cite{Pollack}.  Other attempts have fared equally poorly, making sense 
 for only limited classes of spacetimes or failing to pick out unique constant time hypersurfaces. 

\item{\bf The inner product:}\\[1ex]
In addition to an evolution equation, a quantum theory must provide an inner product on the 
space of states, in order to define probability amplitudes and normalize probabilities.  Here, 
too, the absence of a fixed background time causes trouble.

Let us suppose we have managed to solve the Wheeler-DeWitt equation 
(\ref{time6}), obtaining a space $\{\left|\psi_\alpha[q]\right\rangle\}$ of physical states, which 
for  simplicity we will take to be functions of the spatial metric $q_{ij}$ (the metric
representation).  The metric $q_{ij}$ is defined in (\ref{time1}) at a particular time
$t$, that is, on a particular time slice $\Sigma_t$.  Intuitively, we would like the inner
product $\langle\psi_\alpha|\psi_\beta\rangle$ to refer to wave functions at ``the same
time.''  But the time coordinate $t$ and the slice $\Sigma_t$ have completely dropped out 
of the Wheeler-DeWitt equation, and there is no background spacetime to tell us what 
``the same time'' means.

This problem manifests itself mathematically in the fact that the naive inner product
$$\langle\psi_\alpha|\psi_\beta\rangle = \int[dq]\psi_\alpha^*[q]\psi_\beta[q]$$
is badly divergent.  The divergence comes from the fact that infinitely many spatial 
metrics $q_{ij}$ represent the same physical configuration, defined on time slices
that differ from each other only by coordinate changes.  The Hamiltonian constraint $\widehat H$
generates ``surface deformations'' that move the slice $\Sigma_t$, dragging along the
metric, without changing the actual physics \cite{IshamKuchar}, and the naive inner
product counts all of them.

In principle, we understand how to deal with this problem: we should gauge fix the
inner product, essentially restricting the metric $q_{ij}$ to a single time slice \cite{Woodard}.
In practice, though, this requires finding a universal time slicing, a ``clock'' that makes
physical sense for any spacetime that can appear in a superposition of states.  Once again,
we do not know how to do this.

\item{\bf Operators and causal structure:}\\[1ex]
The third ingredient needed for a quantum theory is a set of observables, self-adjoint
operators that take physical states to physical states.  Here, too, a problem of time
appears.  Suppose $|\psi\rangle$ is a physical state, that is, a state obeying (\ref{time6}), 
and let $\mathcal{\widehat O}$ be an operator.  For $\mathcal{\widehat O}$ to take physical 
states to physical states, we must have ${\widehat H}\mathcal{\widehat O}|\psi\rangle=0$,
which implies in turn that
\begin{equation}
\bigl[\,{\widehat H},\mathcal{\widehat O}\,\bigr]=0 \quad\hbox{on physical states}.
\label{time7}
\end{equation}
Such operators exist, and it may be possible to define them by projection or related constructions
\cite{Chataignier,Chatb,Goeller}.  But they are necessarily nonlocal \cite{Torre,Giddings}, making 
their physical interpretation problematic \cite{Tsamis}.  

This is not merely a technical issue; it goes to the heart of the problem of causal structure in 
quantum gravity.  In standard quantum field theory, the causal structure of spacetime is enforced 
by the condition of ``microcausality'':
\begin{equation}
\bigl[\,\mathcal{\widehat O}_1(x),\mathcal{\widehat O}_2(x')\,\bigr]=0 
    \quad\hbox{if spacetime the points $x$ and $x'$ are spacelike separated}.
\label{time8}
\end{equation}
Such a condition is problematic in a quantum theory of gravity: whether
$x$ and $x'$ are spacelike separated or not depends on the metric, which is no longer
fixed.  The absence of local operators is an explicit manifestation of this problem.

As noted above, (\ref{time7}) also implies that the Heisenberg picture for quantum gravity,
like the Schr{\"o}dinger picture, is ``frozen.''  Indeed, the Heisenberg equation of motion 
for a physical operator is now
\begin{equation}
\frac{d\mathcal{\widehat O}}{dt} = \frac{i}{\hbar}\bigl[\,{\widehat H},\mathcal{\widehat O}\,\bigr]=0 .
\label{time9}
\end{equation}
\end{enumerate}

\section{Covariant phase space \label{cov}}

We now turn to a rather different topic, although one that will ultimately link up with the problem 
of time.  The covariant phase space description of a classical system can be traced
back to Lagrange (see \cite{Bombelli}), but has recently gained traction as a useful way of
obtaining a canonical description while preserving manifest general covariance.  We again 
give a brief review, referring readers to \cite{AshMag,Bombelli,Crn,Crnb,Lee,Waldb,Hu,Carlipb} for 
more comprehensive treatments and technical details.

The standard setting for Hamiltonian mechanics is phase space $\mathcal{M}$,  conventionally 
defined as the space of initial data on a $d$-spatial-dimensional Cauchy surface $\Sigma_t$.  But for a 
theory with a well posed initial value problem, each set of initial data determines a unique classical 
solution, a classical ``history.''  The basic idea of covariant phase space is to reformulate phase 
space as the space of such histories. 

More precisely, in the standard description initial data can be parametrized by canonically conjugate
generalized positions and momenta $(\varphi^i,\pi_i)$, which can together be viewed as coordinates 
$z^A$ on $\mathcal{M}$.  The space $\mathcal{M}$ comes equipped with a symplectic current  
$\omega$, a closed $d$-form (that is, a rank $d$ differential form for which $d\omega=0$),
\begin{equation}
\omega = \delta\pi_i\wedge\delta\varphi^i = \omega_{AB}\delta z^A\wedge \delta z^B ,
\label{cov1}
\end{equation}
where variations $\delta z$ of fields are treated as exterior derivatives on phase space $\mathcal{M}$.
Such a form can be integrated over a $d$-dimensional surface; its integral over a Cauchy surface $\Sigma_t$,
$$\Omega = \int_{\Sigma_t}\omega ,$$
is the symplectic form on $\mathcal{M}$, a two-form on phase space.\footnote{Beware of a potential
source of confusion: $\omega$ is both a $d$-form on $\Sigma_t$ \emph{and} a two-form on the
infinite-dimensional phase space $\mathcal{M}$.}  Up to subtleties concerning boundary terms, the fact 
that $\omega$ is closed ensures that $\Omega$ is independent of the choice of Cauchy surface 
$\Sigma_t$.   

For a theory with gauge symmetries, $\Omega$ is degenerate---gauge variations $\delta_\lambda z$
are null directions---and $\Omega$ is technically a ``presymplectic form.''  This requires a bit more 
work to quotient out the degeneracies \cite{Lee}, a process known as symplectic reduction.  We will 
not need this level of detail here, but there is one subtlety that may be relevant to quantum gravity.  
In order for symplectic reduction to work, the gauge symmetry must act ``nicely'' on the phase 
space, so the quotient space is typically a stratified manifold.  For large classes of solutions of the vacuum 
Einstein equations, this is guaranteed by slice theorems \cite{Ebin,Isenberg}.  But there are known 
truncations of general relativity with certain matter couplings in which these theorems fail \cite{Hoehnb}.  
In such situations, the covariant phase space construction described here may also fail.

Given a symplectic form $\Omega$, we can now write down Poisson brackets between functions 
on phase space,
\begin{equation}
\left\{X,Y\right\} = \int_{\Sigma_t} \frac{\delta X}{\delta z^A}(\omega^{-1})^{AB}\frac{\delta Y}{\delta z^B} .
\label{cov2}
\end{equation}
If we are given a family of transformations---time translations, for instance---of the form $\delta_\tau z^A$,
labeled by a parameter $\tau$, Hamilton's equations of motion become
\begin{equation}
\delta_\tau z^A = (\omega^{-1})^{AB}\frac{\delta H[\tau]}{\delta z^B} ,
\label{cov3}
\end{equation}
which can be taken as a definition of the Hamiltonian for the transformation.  Slightly more obscurely,
(\ref{cov3}) can be written as
\begin{equation}
\delta H[\tau] = \Omega[\delta z,\delta_\tau z] ,
\label{cov4}
\end{equation}
the defining equation for the Hamiltonian.  

Suppose now that we are dealing with a system with a well posed initial value problem, that is, a 
system for which any set of initial data $(\varphi^i,\pi_i)$ determines a unique solution.  Then each
point in the phase space $\mathcal{M}$ defined on a Cauchy surface $\Sigma_t$ determines a
classical solution $\Phi$.  Conversely, given a Cauchy surface $\Sigma_t$, we can restrict any
classical solution to that surface to determine a point in $\mathcal{M}$.  This provides a one-to-one
map between $\mathcal{M}$ and the space ${\overline{\mathcal{M}}}$ of classical solutions, or
histories.  Indeed, consider the map
\begin{equation}
\tau_{\scriptscriptstyle{\Sigma_t}}: {\overline{\mathcal{M}}} \rightarrow \mathcal{M} \qquad
\Phi \mapsto (\varphi^i,\pi_i)\left|_{\Sigma_t}\right.
\label{cov5}
\end{equation}
that takes any classical solution to its initial data on the slice $\Sigma_t$.  The well-posedness
of the initial value problem tells us that this map is invertible, and requiring it to be a diffeomorphism 
determines a topology on ${\overline{\mathcal{M}}}$ (although with some mathematical subtleties; 
see, for instance, chap.\ 2 of \cite{Woodhouse}).

The space of classical solutions has a natural symplectic form $\overline\Omega$, defined entirely 
from the variation of the spacetime action \cite{Crn,Crnb,Lee,Waldb}.  We will not need details 
here; see Appendix A of \cite{Carlipb} for a short review.  Moreover, the map (\ref{cov5}) 
is a symplectomorphism, taking $\overline\Omega$ to $\Omega$ and thus preserving the
phase space structure.   The pair $({\overline{\mathcal{M}}},{\overline\Omega})$ is commonly known 
as ``covariant phase space,'' covariant in the sense that it requires no choice of a time slice $\Sigma_t$.

For a large class of diffeomorphism invariant theories, including general relativity, one can find an
explicit expression for a Hamiltonian $\overline{H}$, obeying (\ref{cov4}), that generates diffeomorphisms 
\cite{Lee,Waldb}.  This $\overline{H}$ is the covariant phase space version of the ADM Hamiltonian 
(\ref{time5a}), and is again a constraint, vanishing (up to possible boundary terms) on physical solutions.

Note that while $({\overline{\mathcal{M}}},{\overline\Omega})$ and $(\mathcal{M},\Omega)$ are
isomorphic, they are not \emph{canonically} isomorphic.  The isomorphism depends on the choice
of Cauchy surface $\Sigma_t$, and in general there is no preferred choice.  This has two implications 
\cite{Hu,Woodhouse}:
\begin{enumerate}
\item Points in the conventional phase space $\mathcal{M}$ can be thought of as coordinates
for ${\overline{\mathcal{M}}}$.  While it is sometimes possible to find a global parametrization of
classical solutions, for instance in terms of constants of motion, this is often difficult.  But initial data 
on a fixed time slice provide a useful set of labels, and we can interpret $\tau_{\scriptscriptstyle{\Sigma_t}}$ 
as a coordinate map.
\item Consider a family $\Sigma_{\{t\}}$ of Cauchy surfaces.  Suppose we are given initial data
$(\varphi^i,\pi_i)$ on $\Sigma_t$ and want to the fields at a later slice $\Sigma_{t'}$.  From (\ref{cov5}),  
\begin{equation}
(\varphi^i,\pi_i)\left|_{\Sigma_{t'}}\right. 
   = \tau_{\scriptscriptstyle{\Sigma_{t'}}} \circ  \tau^{-1}_{\scriptscriptstyle{\Sigma_t}}%
   (\varphi^i,\pi_i)\left|_{\Sigma_t}\right. 
\label{cov6}
\end{equation}
But this is simply a change of coordinates on ${\overline{\mathcal{M}}}$.  We can thus view time
evolution as a coordinate change on covariant phase space.

Note that we have made no assumptions about $\Sigma_t$ and $\Sigma_{t'}$ except that they
are Cauchy surfaces.  This does not, of course, absolve us from solving any equations---the
covariant phase space approach implicitly assumes that we know the full classical solutions.
But it does provide us with a Hamiltonian description of evolution that requires no preferred
choice of time slicing.
\end{enumerate}

We can thus treat phase space either as a space of initial data or as a space of 
classical histories.  Mathematically the two descriptions are isomorphic.  Physically, however, 
they give very different pictures.  The ordinary phase space picture is one of evolution in time:
one starts with initial data at a fixed time and determines its future development.
The covariant phase space picture is inherently time independent: its elements are entire
histories, and ``time'' is merely a coordinate choice.  There is nothing in the physics that
prefers one view over the other.  But the covariant phase space formalism is perhaps a bit
broader in its reach, requiring a space of solutions but not necessarily a good initial value
problem.

\section{Covariant canonical quantization \label{quant}}

The question is now whether this classical picture of covariant phase space can be used to build 
a quantum theory.  For simple enough cases---for instance, free scalar fields in curved spacetime 
\cite{AshMag}, (2+1)-dimensional gravity with simple topologies 
\cite{Carlip_time,Carlip_comp,Carlip_dirac,Carlip_mod,Carlip_mod2,Kraus}, 
some models of two-dimensional gravity \cite{Navarro}, and certain truncations of general relativity 
\cite{Hu}---this is known to be possible.  For more complicated theories, such as the full
general theory of relativity, it is less clear.  For instance, without a general solution of the field
equations we cannot fully characterize the covariant phase space, although there may be
sensible ways to use classical perturbation theory \cite{Basu}.

We will argue here that if such a quantization is possible, it will provide a major step towards
solving the difficulties described in \S\ref{time}.  While the problem of time would not be
entirely eliminated, it would be reduced to the more tractable problem of time in
classical general relativity.  

Let us suppose we have found, at least locally, a canonical set of coordinates 
$\{{\bar\varphi}^a,{\bar\pi}_a\}$ on covariant phase space ${\overline{\mathcal{M}}}$, that is, a 
set for which the symplectic form is ${\overline\Omega} = \delta{\bar\pi}_a\wedge\delta{\bar\varphi}^a$.
Typically, ${\overline{\mathcal{M}}}$ is infinite dimensional; the index $a$ corresponds roughly to
the pair $(i,x)$ in the standard canonical formulation.  For simplicity, we are  taking this index
to be discrete---for instance, labeling coefficients of a basis of functions---but a generalization should 
be possible.  We will assume the simplest form of quantization, in which the Poisson brackets 
of these coordinates become commutators,
\begin{align}
\left[\hat{\bar\varphi}^a,\hat{\bar\pi}_b\right] &= i\hbar\,\delta^a_b , \nonumber\\
\left[\hat{\bar\varphi}^a,\hat{\bar\varphi}^b\right] &= \left[\hat{\bar\pi}_a,\hat{\bar\pi}_b\right] = 0 ,
\label{quant1}
\end{align}
and we will take our wave functions $\left|\Psi[{\bar\varphi}]\right\rangle$ to be in the position 
representation, that is, functions of the generalized positions $\bar\varphi$.  
There are, of course, other possibilities---one might choose different fundamental variables and
commutators, for instance \cite{Ishamb}, or use a more sophisticated technique such as 
deformation quantization \cite{deform}---but these are universal issues in quantization, not particular 
to covariant phase space.  As in \S\ref{time}, we must also provide an inner product on the space
$\{\left|\Psi[{\bar\varphi}]\right\rangle\}$ of wave functions, which will normally require gauge fixing the naive inner 
product.  Some ideas of how to do this have been proposed \cite{Woodard,Chataignier,Chatb,Marolf},
but for quantum gravity the problem remains very difficult; we will assume it has been solved,
at least in some systematic method of approximation.

If our theory is diffeomorphism invariant, the classical Hamiltonian $\overline{H}$ of \S\ref{cov} 
will still be a constraint, and instead of a Schr{\"o}dinger equation, a Wheeler-DeWitt equation
\begin{equation}
{\widehat{\overline{H}}}\left|\Psi[\bar\varphi]\right\rangle = 0
\label{quant2}
\end{equation}
will again hold.  Now, however, this condition has a simple interpretation: as functions on the
space of histories, wave functions determine amplitudes for entire block universes,
and are thus inherently time independent.   The question is whether we can now obtain a picture 
that exhibits time evolution.

To do so, start with the classical description of ${\overline{\mathcal{M}}}$, and pick a Cauchy
surface $\Sigma_t$.  Up to ordering ambiguities, the map $\tau_{\scriptscriptstyle{\Sigma_{t}}}$
of (\ref{cov5}) translates to a map between operators in the quantum theory,
\begin{equation}
{\hat\tau}_{\scriptscriptstyle{\Sigma_{t}}}: 
   ({\hat{\bar\varphi}}^a,{\hat{\bar\pi}}_a) \mapsto ({\hat\varphi}^i(\Sigma_t) ,{\hat\pi}_i(\Sigma_t)) .
\label{quant3}
\end{equation}
This gives us a set of slice-dependent operators $\{{\hat\varphi}^i(\Sigma_t) ,{\hat\pi}_i(\Sigma_t)\}$,
which depend  on the covariant phase space operators $\{\hat{\bar\varphi}^a,\hat{\bar\pi}_a\}$ and
the \emph{classical} data that determine the map ${\tau}_{\scriptscriptstyle{\Sigma_{t}}}$.  These
are examples of what Rovelli has called ``evolving constants of motion''  \cite{Rovellib}, operators
that commute with the generator of time translations but depend on a ``time'' parameter---here,
the slice $\Sigma_t$---that is determined by the classical dynamics.

From the canonical commutation relations (\ref{quant1}) and the fact that ${\tau}_{\scriptscriptstyle{\Sigma_{t}}}$
is a symplectomorphism, it follows that the operators ${\hat\varphi}^i(\Sigma_t)$
commute with each other, and  should at least formally comprise a maximal set of commuting operators. 
Diagonalizing them thus gives us a new ``evolving basis,''
\begin{equation}
{\hat\varphi}^i(\Sigma_t)|{\varphi}^i(\Sigma_t)\rangle = {\varphi}^i(\Sigma_t)|{\varphi}^i(\Sigma_t)\rangle ,
\label{quant4}
\end{equation}
which depends on the classical choice of $\Sigma_t$.  Any covariant canonical state $|\Psi[\bar\varphi]\rangle$ 
can be expanded in this basis, and the components 
$$\langle {\varphi}^i(\Sigma_t) |\Psi[\bar\varphi]\rangle$$
are ``time'' dependent, depending on the classical slice $\Sigma_t$.  Moreover, if we choose a
family $\Sigma_{\{t\}}$ of slices that foliate some region of classical spacetime, completeness of the 
bases $|{\varphi}^i(\Sigma_t)\rangle$ and $|{\varphi}^i(\Sigma_{t'})\rangle$ implies that the functions 
$\langle {\varphi}^i(\Sigma_t) |\Psi[\bar\varphi]\rangle$ and $\langle {\varphi}^i(\Sigma_{t'}) |\Psi[\bar\varphi]\rangle$ 
are related by a formally unitary transformation. For any particular choice of time slicing, we can therefore 
write
\begin{equation}
i\hbar\frac{d\ }{dt} \langle {\varphi}^i(\Sigma_t) |\Psi[\bar\varphi]\rangle 
   = {\widehat{\mathcal{H}}}(\Sigma_{\{t\}})\langle {\varphi}^i(\Sigma_t) |\Psi[\bar\varphi]\rangle
\label{quant5}
\end{equation}
for some Hermitian operator ${\widehat{\mathcal{H}}}(\Sigma_{\{t\}})$.  Note that ${\widehat{\mathcal{H}}}$ 
is \emph{not} the Hamiltonian constraint of (\ref{quant2}), and that it will generally differ for different 
choices of the family $\Sigma_{\{t\}}$.  For any such family, the ${\widehat{\mathcal{H}}}(\Sigma_{\{t\}})$ 
are again evolving constants of motion, defined on the covariant phase space but depending on
a classical choice of foliation.

Now consider an initial slice $\Sigma_i$ and a final slice $\Sigma_f$.  There are infinitely many
ways to foliate the spacetime between them, each with its own ${\widehat{\mathcal{H}}}(\Sigma_{\{t\}})$.
This proliferation of Hamiltonians leads to the ``multiple choice problem'' of time \cite{Kuchar,Isham}:
one must worry whether the ultimate evolution from $\Sigma_i$ to $\Sigma_f$ depends on one's
intermediate choices.  Here, though, at least in principle this is not a problem: by construction,
the final wave function $\langle {\varphi}^i(\Sigma_f) |\Psi[\bar\varphi]\rangle$ depends only on the
covariant phase space state $|\Psi[\bar\varphi]\rangle$ and the final slice.

As we cautioned in the introduction, these argument are not mathematically rigorous.  For
simple examples like those of \cite{Hu}, the constructions can be carried out in full.  But for
more realistic cases, many technical problems remain.  We have assumed that an inner
product on the covariant Hilbert space has been found; for gravity this is an unsolved problem.   
The passage (\ref{quant3}) from functions to operators is plagued by operator ordering ambiguities.
Arguments about completeness of bases and unitarity become much more delicate for operators
with continuous spectra.  Results that are clear for finite dimensional Hilbert spaces do not
always extend easily; even in a system as simple as a free scalar field in more than two
dimensions, infinite sums involved in basis changes can reintroduce a multiple choice problem
\cite{Varadarajan}.  

For quantum gravity, a first attempt at a rigorous analysis was made in \cite{Cosgrove}, but work
since then has largely focused on simpler models.  One might hope, though, that the remaining
problems, while difficult, are in some sense ``technical,'' and that the general conceptual 
approach to the problem of time will survive and more detailed mathematical analysis.

\section{An example from (2+1)-dimensional gravity \label{example}}

We now turn to a specific application of this formalism, quantum gravity for a spatially compact
universe in 2+1 dimensions, that is, two dimensions of space plus one of time.  It has been 
understood for 50 years \cite{Star} that such a reduction in dimension vastly simplified general
relativity (see \cite{Carlip_rev} for a review).  In 2+1 dimensions the vacuum Einstein equations 
imply that spacetime is flat---or constant curvature if the cosmological constant is nonzero---and
thus contains no propagating degrees of freedom.\footnote{A (2+1)-dimensional spacetime with 
an asymptotic boundary may have ``boundary gravitons'' that propagate by the boundary Hamiltonian,
but as noted earlier, we are not dealing with that case here.}  But while this drastically simplifies
the theory, it does not make it completely trivial: a topologically nontrivial spacetime will still carry
global geometric degrees of freedom, which introduce the same conceptual problems we face in
realistic (3+1)-dimensional gravity.

\begin{figure}
\centerline{%
\begin{picture}(200,100)(0,0)
\put(10,0){\line(1,0){210}}
\put(10,0){\line(0,1){110}}
\thicklines
\put(10,0){\line(1,2){40}}
\put(160,0){\line(1,2){40}}
\put(10,0){\line(1,0){150}}
\put(50,80){\line(1,0){150}}
\put(7,-10){$\scriptstyle 0$}
\put(157,-10){$\scriptstyle 1$}
\put(49,83){$\scriptstyle\tau$}
\put(200,83){$\scriptstyle\tau+1$}
\put(224,-4){$\scriptstyle x$}
\put(12,110){$\scriptstyle y$}
\end{picture}
}
\caption{A flat torus may be formed by ``gluing'' opposite edges.  The complex modulus $\tau$ determines 
the global geometry; in general, tori with different moduli are not diffeomorphic.}
\label{fig1}
\end{figure}
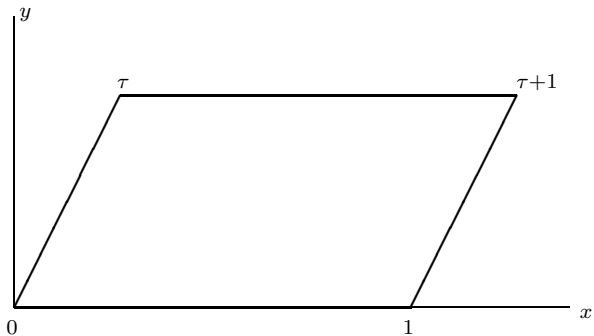
 
As a warmup, consider a flat two-dimensional torus $T^2$.  Such a space can always be constructed
by cutting out a parallelogram from the plane and identifying opposite edges.  After rescaling one
circumference to $1$, distinct tori are parametrized by a single complex number $\tau$, the
modulus, as shown in figure \ref{fig1}.  In the coordinates shown in the figure, the metric is simply
\begin{equation}
ds^2 = \left|dx + \tau dy\right|^2 ,
\label{example1}
\end{equation}
with edges identified by translations $x\sim x+1$ and $y\sim y+1$.  These identifications are
isometries of the Euclidean plane, as they must be to maintain flatness.  The metric (\ref{example1})
is manifestly flat, but in general there is no diffeomorphism that will take a metric with one modulus
$\tau$ to a metric with another: the modulus is a genuine geometric degree of freedom.

Now consider a flat (2+1)-dimensional spacetime with topology $\mathbb{R}\times T^2$, that is,
a spatially closed three-dimensional torus universe.  Just as a two-dimensional torus can be obtained 
from a region of the plane by isometrically identifying the edges, any flat $\mathbb{R}\times T^2$
can be obtained from a region of flat (2+1)-dimensional Minkowski space by isometrically identifying 
the edges.  It can be shown that the allowable isometries take the form 
\cite{Carlip_time,Carlip_comp,Carlip_dirac,Carlip_mod,Carlip_mod2}
\begin{align}
&\Lambda_1: (t,x,y)\rightarrow (t\cosh\lambda + x\sinh\lambda, x\cosh\lambda+t\sinh\lambda,y+a)
\nonumber\\
&\Lambda_2: (t,x,y)\rightarrow (t\cosh\mu + x\sinh\mu, x\cosh\mu+t\sinh\mu,y+b) ,
\label{example2}
\end{align}
now labeled by four real parameters $(\mu,\lambda,a,b)$.  These parameters act as 
coordinates for the covariant phase space, uniquely labeling classical solutions with the specified
 topology.  Moreover, they come with a natural symplectic structure $\overline{\Omega}$, 
 obtained as in \S\ref{cov} and mathematically related to the symplectic structure of loops on 
 a surface \cite{Goldman}.  The quantization described in \S\ref{quant} is then straightforward: 
 the nonvanishing commutators are
\begin{equation}
[{\hat\mu},{\hat a}] = [{\hat b},{\hat\lambda}] = -\frac{i\hbar}{2} ,
\label{example3}
\end{equation}
and we can thus take our wave functions to be functions $\Psi(\lambda,\mu)$, with
\begin{equation}
{\hat a} = \frac{i\hbar}{2}\frac{\partial\,}{\partial\mu}, \qquad 
{\hat b} = -\frac{i\hbar}{2}\frac{\partial\,}{\partial\lambda} .
\label{example4}
\end{equation}

The wave functions $\Psi(\lambda,\mu)$ are time independent, and give amplitudes for entire
block universes.  To obtain time-dependent operators, we follow the procedure of \S\ref{quant},
starting with a choice of classical time slicing.  Here there is a natural choice; in this (2+1)-dimensional
setting, York's extrinsic time always exists, and gives a unique slicing $K=-t$ \cite{Andersson}.
Moreover, the slices of constant time $t$ are spatially flat.  With this choice, space at a fixed time 
is thus a flat torus, characterized by a modulus $\tau(t)$
and its conjugate momentum $p(t)$.  By looking at the explicit form of the spacetime metric,
we can obtain this quantities in terms of the parameters $(\mu,\lambda,a,b)$, both classically
and quantum mechanically:
\begin{equation}
{\hat\tau}(t) = \left({\hat a} + \frac{i\hat\lambda}{t}\right)^{-1}%
  \left({\hat b} + \frac{i\hat\mu}{t}\right),\quad
{\hat p}(t) =-it\left({\hat a} - \frac{i{\hat\lambda}}{t}\right)^2  .
\label{example5}
\end{equation}

As in \S\ref{quant}, we can now diagonalize the operators $\hat\tau$ and ${\hat\tau}^\dagger$, 
finding eigenstates for which $\hat\tau|\psi\rangle = \tau|\psi\rangle$ and
${\hat\tau}^\dagger|\psi\rangle = {\bar\tau}|\psi\rangle$.  A straightforward calculation 
\cite{Carlip_time,Carlip_comp,Carlip_dirac,Carlip_mod,Carlip_mod2} yields
\begin{equation}
\langle \lambda,\mu|\tau,{\bar\tau},t\rangle = K(\lambda,\mu|\tau,{\bar\tau},t)
   = \left(\frac{\mu - \tau\lambda}{\tau_2^{1/2}t}\right)\exp\left\{\frac{-i|\mu-\tau\lambda|^2}{\tau_2t}\right\} .
\label{example6}
\end{equation}
The slice-dependent wave functions  
$\Psi(\tau,{\bar\tau},t) = \int\!d\lambda d\mu\, K(\lambda,\mu|\tau,{\bar\tau},t)\Psi(\lambda,\mu)$
can then be showed to satisfy a Schr{\"o}dinger equation
\begin{equation}
i\frac{d\,}{dt}\Psi(\tau,{\bar\tau},t) = {\widehat{\mathcal{H}}}\Psi(\tau,{\bar\tau},t) \quad
\hbox{with\quad ${\widehat{\mathcal{H}}} = t^{-1}\left(\Delta_{1/2}\right)^{1/2}$} ,
\label{example7}
\end{equation}
where $\Delta_{1/2}$ is a Laplacian on moduli space.

This essentially reproduces the result one would obtain by first fixing the York time slicing and then
quantizing \cite{Carlip_time}.  We say ``essentially'' because the operator ordering in (\ref{example5})
is not unique, and different choices lead to $\mathcal{O}(\hbar)$ modifications of the Hamiltonian.
For (2+1)-dimensional gravity, though, an additional symmetry reduces the ambiguity.  The mapping class
group---the group of ``large'' diffeomorphisms, diffeomorphisms that cannot be continuously deformed
to the identity---acts nontrivially on the space of moduli $\tau$, and the wave function should
transform under a representation of this group.  It turns out \cite{Carlip_dirac,Carlip_mod,Carlip_mod2}
that the choice of representation completely determines the operator ordering, the inner product, the
normalization of the wave functions (\ref{example6}), and the Hamiltonian, all of the ingredients needed 
for a sensible quantum theory.

We thus have an explicit example in which a covariant phase space quantum theory of gravity,
with no intrinsic time dependence, can be transformed into a ``time''-dependent Schr{\"o}dinger
picture.  This Schr{\"o}dinger gives a physically sensible quantum theory: for instance, the wave
functions (\ref{example6}) are peaked on the classical trajectories \cite{Ezawa}.  But we have found
the time dependence using only classical properties, in a way that would clearly work as well for
any other classical choice of time slicing.

\section{Conclusions}

Covariant canonical quantization does not solve the problem of time in quantum gravity.  It
does, however, recast some of the questions in a more tractable form.  In particular, one 
need not pick a preferred time slicing, classically or quantum mechanically.  Rather, at least 
in principle, any classical choice of time slicing provides a set of slice-dependent
operators that determine a Schr{\"o}dinger picture time evolution for that foliation.  

The approach also offers a physicists' perspective on the ``block universe/evolving universe''
question, although perhaps not one that addresses all of the philosophical issues.  In
covariant canonical quantization, there are an infinite number of descriptions of an evolving
universe, one for each time slicing.  But these are all isomorphic to a timeless block universe 
description coming from the quantization of the covariant phase space, and one's choice of 
which description to use seems to have no physical content.  Classically, this isomorphism 
requires that the systems under consideration be deterministic---that is the 
significance of a well posed initial value problem---but it is hard to see how to do physics 
without such an assumption.  Quantum mechanically the equivalence is much less clear in
other formulations, but here it is automatic.

All this comes with a proviso: ``assuming the program can be carried out.''  We do
not wish to understate the potential problems.  To fully define the covariant phase space, we
need a complete characterization of the classical solutions, although we may be able to
make progress with classical perturbation theory \cite{Basu}.  In the example of \S\ref{example}
we had a diffeomorphism-invariant parametrization of the space of solutions, but if we do 
not---if, for instance, we use initial data as coordinates as in \S\ref{cov}---we will need to
gauge fix the inner product \cite{Woodard}, a process easy to describe but difficult  to implement.  
To obtain a time-dependent picture we need a classical time slicing; an infinite number of
such slicings exist, but we do not know to explicitly describe even one in a way that holds
for the entire space of solutions.  Physical operators on a slice will be nonlocal (as the moduli
$\tau$ of \S\ref{example} were), and hard to construct.  And as in \S\ref{example}, we expect
serious operator ordering ambiguities in the Hamiltonian, though perhaps these will be limited
by the demand for consistent evolution \cite{Cosgrove}.
 
If we are lucky, though, these problems are ``technical,'' requiring hard work but no radical
new ideas.  One could imagine, for instance, a systematic classical perturbative expansion
of the field equations---the post-Minkowskian expansion, for instance \cite{Damour}---combined
with an order by order specification of an inner product and spatial slicing.  Whether this
is possible remains to be seen.  But if it is, then perhaps some of the conceptual
problems of quantum gravity are not quite as severe as we have believed.
 
\vspace*{.5ex}
\begin{flushleft}
\large\bf Acknowledgments
\end{flushleft}

We would like to thank Cliff Taubes for catching an error in an earlier version of
this paper.  This work was supported in part by Department of Energy grant
DE-FG02-91ER40674.

\end{document}